# Properties of convergence of nonextensive statistical distribution to the Lévy distribution

Sumiyoshi Abe[1] and A. K. Rajagopal[2]

[1]*College of Science and Technology, Nihon University,*

*Funabashi, Chiba 274-8501, Japan*

[2]*Naval Research Laboratory, Washington, DC 20375-5320*

It is shown that the distribution derived from the principle of maximum Tsallis entropy is a superposable Lévy-stable distribution. Concomitantly, the leading order correction to the limit distribution is also deduced. This demonstration fills an important gap in the derivation of the Lévy-stable distribution from the nonextensive statistical framework.

PACS numbers: 05.40.Fb, 05.40.-a, 05.20.-y

Recently, Prato and Tsallis [1] gave an argument to demonstrate the appearance of the Lévy-like distribution function from the nonextensive statistical framework applied to the problems of random walk and diffusion phenomena. The distribution they obtained has the same asymptotic behavior as that of the Lévy distribution for large values of the random variable. Then, employing the Lévy-Gnedenko generalized central limit theorem [2,3], they claim that their distribution approaches the exact Lévy distribution by convolution of the distribution many times. However, an important question remains regarding the convergence of the convolution. The purpose of this paper is to examine this question and to explore the rate of convergence.

In Ref. [1], the authors consider the maximization of the Tsallis entropy [4]

$$S_q[p] = \frac{1}{1-q}\left\{\int_{-\infty}^{\infty} \frac{dx}{\sigma}[\sigma p(x)]^q - 1\right\},\qquad(1)$$

under the constraints on the normalization condition of the distribution

$$\int_{-\infty}^{\infty} dx\, p(x) = 1,\qquad(2)$$

and on the generalized second moment [5]

$$\int_{-\infty}^{\infty} dx\, x^2\, P(x) = \sigma^2,\qquad(3)$$

where $\sigma$ is a positive constant giving a length scale and $P(x)$ is the escort distribution associated with $p(x)$ [6]

$$P(x) = \frac{[p(x)]^q}{\int_{-\infty}^{\infty} dx'\,[p(x')]^q}.\qquad(4)$$

$q$ in eqs. (1) and (4) is a parameter, which is assumed here to be

$$\frac{5}{3} < q < 3. \tag{5}$$

The maximum-entropy distribution is found to be

$$p(x) = \frac{1}{\sigma}\sqrt{\frac{q-1}{\pi(3-q)}} \frac{\Gamma\left(\frac{1}{q-1}\right)}{\Gamma\left(\frac{3-q}{2q-2}\right)} \frac{1}{\left(1+\frac{q-1}{3-q}\frac{x^2}{\sigma^2}\right)^{1/(q-1)}}. \tag{6}$$

The range in eq. (5) covers the same range as that of the Lévy distribution. [See eq. (10) below.] It should be noted that the ordinary second moment with respect to this distribution is divergent but the generalized second moment in eq. (3) is finite. For large values of $x$, the above distribution asymptotically behaves as

$$p(x) \sim x^{-2/(q-1)}. \tag{7}$$

Comparing this with the symmetric Lévy distribution

$$L_\gamma(x) = \frac{1}{2\pi} \int_{-\infty}^{\infty} dk \, \exp(-ikx) \exp(-a|k|^\gamma) \quad (0 < \gamma < 2) \tag{8}$$

with a positive constant $a$, whose asymptotic behavior is

$$L_\gamma(x) \sim x^{-1-\gamma}, \tag{9}$$

we find that $q$ is related to the Lévy index $\gamma$ as follows:

$$q = \frac{3+\gamma}{1+\gamma}. \tag{10}$$

The generalized central limit theorem is stated as follows. Given a distribution whose ordinary second moment is divergent, its $N$-fold convolution approaches one of a stable class of distributions $\{L_\gamma(x): 0 < \gamma < 2\}$ for large $N$ provided that the convolution converges. This property has not been ascertained in Ref. [1]. In what follows, we discuss this problem using the method of characteristic functions.

In order to apply the generalized central limit theorem, it is necessary to consider the sum of $N$ scaled independent random variables $\{x_i\}_{i=1,2,\ldots,N}$ [2,3]:

$$x = \frac{x_1 + x_2 + \ldots + x_N}{B_N}. \tag{11}$$

The scaling factor $B_N$ here has to be chosen in such a way that the limit distribution is independent of the number $N$ of convolutions. Suppose each $x_i$ obey the Lévy distribution in eq. (8). Then, the distribution of $x$ is given by

$$L_\gamma^{(N)}(x) = B_N \left( L_\gamma * \ldots * L_\gamma \right)(B_N x). \tag{12}$$

The characteristic function of $L_\gamma^{(N)}(x)$ is calculated to be

$$\chi_L^{(N)}(k) = \int_{-\infty}^{\infty} dx \, \exp(ikx) \, L_\gamma^{(N)}(x)$$

$$= \left[ \chi_L\left(\frac{k}{B_N}\right) \right]^N, \tag{13}$$

where $\chi_L(k)$ is the characteristic function of $L_\gamma(x)$:

$$\chi_L(k) = \exp\left(-a|k|^\gamma\right). \tag{14}$$

Therefore, we have

$$B_N = N^{1/\gamma}, \tag{15}$$

where the constant of proportionality is set equal to unity for the sake of simplicity.

To examine the convergence properties of the distribution in eq. (6), first we calculate its characteristic function. After some algebra, we find

$$\chi(k) = \int_{-\infty}^{\infty} dx \, \exp(ikx) \, p(x)$$

$$= 2^{1-v} \frac{(\lambda |k|)^v}{\Gamma(v)} K_v(\lambda |k|), \tag{16}$$

where $\Gamma(z)$ and $K_v(z)$ are the Euler gamma function and the modified Bessel function [7], respectively, and

$$v = \frac{3-q}{2q-2} = \frac{\gamma}{2}, \tag{17}$$

$$\lambda = \sqrt{2v}\sigma. \tag{18}$$

Clearly, the range of $v$ is

$$0 < v < 1. \tag{19}$$

Now, let us discuss the limit $N \to \infty$ of

$$f(k; N) \equiv \left[\chi\left(\frac{k}{N^{1/\gamma}}\right)\right]^N \tag{20}$$

in order to determine the convergence properties of the distribution in eq. (6). For this purpose, we take the logarithm of $f(k; N)$:

$$\ln f(k; N) = N \ln \left[ 2^{1-\gamma/2} \frac{\left(\frac{\lambda|k|}{N^{1/\gamma}}\right)^{\gamma/2}}{\Gamma\left(\frac{\gamma}{2}\right)} K_{\gamma/2}\left(\frac{\lambda|k|}{N^{1/\gamma}}\right) \right]. \quad (21)$$

Using the formula $K_\nu(z) = \frac{\pi}{2} \frac{I_{-\nu}(z) - I_\nu(z)}{\sin(\nu\pi)}$ [7], where $I_\nu(z)$ is essentially the Bessel function of imaginary argument, and evaluating the series expansion $I_\nu(z) = \sum_{k=0}^{\infty} \frac{(z/2)^{2k+\nu}}{k!\Gamma(k+\nu+1)}$ [7], we find

$$\ln f(k; N) = N \ln \left[ 1 - \frac{1}{N} \frac{\Gamma\left(1-\frac{\gamma}{2}\right)}{\Gamma\left(1+\frac{\gamma}{2}\right)} \left(\frac{\lambda|k|}{2}\right)^\gamma + \frac{2}{2-\gamma} \frac{1}{N^{2/\gamma}} \left(\frac{\lambda|k|}{2}\right)^2 + \text{L} \right]$$

$$= -\frac{\Gamma\left(1-\frac{\gamma}{2}\right)}{\Gamma\left(1+\frac{\gamma}{2}\right)} \left(\frac{\lambda|k|}{2}\right)^\gamma + \frac{2}{2-\gamma} \frac{1}{N^{2/\gamma-1}} \left(\frac{\lambda|k|}{2}\right)^2$$

$$- \frac{1}{2N} \left[\frac{\Gamma\left(1-\frac{\gamma}{2}\right)}{\Gamma\left(1+\frac{\gamma}{2}\right)}\right]^2 \left(\frac{\lambda|k|}{2}\right)^{2\gamma} + \text{L} . \quad (22)$$

Therefore, in the limit $N \to \infty$, we obtain the exact Lévy distribution with

$$a = \frac{\Gamma\left(1-\frac{\gamma}{2}\right)}{\Gamma\left(1+\frac{\gamma}{2}\right)} \left(\frac{\lambda}{2}\right)^\gamma. \quad (23)$$

Simultaneously, we also find that dominant correction to the Lévy distribution is of $O(N^{-1})$ if $0 < \gamma < 1$ whereas it is of $O(N^{1-2/\gamma})$ if $1 < \gamma < 2$.

In conclusion, we have shown that the distribution derived from the principle of maximum Tsallis entropy is in fact a superposable Lévy-stable distribution. We have also determined the leading order correction to this limit distribution. We have thus filled an important gap in the derivation of the Lévy-stable distribution from the nonextensive statistical approach.

One of us (S.A.) was supported by the GAKUJUTSU-SHO Program of College of Science and Technology, Nihon University. He thanks the warm hospitality of the Naval Research Laboratory, Washington, DC, which made this collaboration possible. The other author (A. K. R.) acknowledges the partial support from the US Office of Naval Research.